\documentclass[lettersize,journal]{IEEEtran}
\usepackage{amsmath,amsfonts}

\usepackage{textcomp}
\usepackage{stfloats}
\usepackage{url}
\usepackage{verbatim}
\usepackage{graphicx}
\usepackage{cite}
\usepackage{xcolor}
\usepackage{subfigure} 
\usepackage{multirow}
\usepackage{enumitem}  
\usepackage{cuted}     
\usepackage{amssymb}
\usepackage{adjustbox}
\usepackage{booktabs}
\usepackage{array}

\usepackage[ruled,vlined,linesnumbered]{algorithm2e}
\begin{document}

\title{Deep Learning-Based Site-Specific Channel Modeling and Inference}

\author{Junzhe Song,
Ruisi He,~\IEEEmembership{Senior Member,~IEEE,}
Mi Yang,~\IEEEmembership{Member,~IEEE,}\\
Zhengyu Zhang,~\IEEEmembership{Student Member,~IEEE,}
Shuaiqi Gao,
Bo Ai,~\IEEEmembership{Fellow,~IEEE,}
Zhangdui Zhong, ~\IEEEmembership{Senior Member, IEEE}

\thanks{
J. Song, R. He, M. Yang, Z. Zhang, S. Gao, and B. Ai are with the Schooof Electronics and Information Engineering, Beijing Jiaotong University, Beijing 100044, China (email: 25115063@bjtu.edu.cn; ruisi.he@bjtu.edu.cn; myang@bjtu.edu.cn; 21111040@bjtu.edu.cn; 25110046@bjtu.edu.cn; boai@bjtu.edu.cn; zhdzhong@bjtu.edu.cn).



}}

\maketitle

\begin{abstract}
Site-specific channel inference plays a critical role in the design and evaluation of next-generation wireless communication systems by considering the surrounding propagation environment. However, traditional methods are unscalable. Recently, satellite imagery has emerged as a valuable modality containing rich propagation information for AI-based channel prediction. However, existing approaches using these images are limited to predicting large-scale fading parameters, lacking the capacity to reconstruct the complete channel impulse response (CIR). To address this limitation, we propose a deep learning-based site-specific channel modeling and inference framework using satellite images to predict structured Tapped Delay Line (TDL) parameters. We first establish a joint channel-satellite dataset based on measurements. Then, a novel deep learning network is developed to reconstruct the channel parameters. Specifically, a cross-attention-fused dual-branch pipeline extracts macroscopic and microscopic environmental features, while a recurrent tracking module captures the long-term dynamic evolution of multipath components. Experimental results demonstrate that the proposed method achieves high-quality reconstruction of the CIR in unseen scenarios, with a Power Delay Profile (PDP) Average Cosine Similarity exceeding 0.96. This work provides a pathway toward site-specific channel inference for future dynamic wireless networks.
\end{abstract}


\IEEEdisplaynontitleabstractindextext

\IEEEpeerreviewmaketitle
\section{Introduction}
\IEEEPARstart{S}{ite-specific} channel models characterize communication links by considering the surrounding environmental geometry (e.g., buildings, vegetation, and moving objects) \cite{Zemen-p1site}. With the advancement of 5G and the evolution towards 6G, wireless scenarios are evolving towards high mobility and dynamism (e.g., autonomous driving, vehicular communications \cite{8851421}, and unmanned aerial vehicle (UAV) networks) \cite{vardhan2025aber,tian2025analytical}. The propagation characteristics in these scenarios are complex, and they are affected simultaneously by large-scale fading and rich multipath effects caused by dense scatterers \cite{he2024wireless}. Therefore, obtaining high-fidelity Channel Impulse Response (CIR) data for these specific sites is fundamental to the design and performance evaluation of next-generation communication systems \cite{he2015characterization,zhang2025non}.

Existing standardized models such as 3GPP channel models \cite{zhu20213gpp}, WINNER II \cite{kyosti2007winner}, COST \cite{liu2012cost,he2025cost}, and the map-based METIS model\cite{unknown} (e.g., the Tapped Delay Line (TDL) models defined in 3GPP TR 38.901 \cite{zhu20213gpp}, which parameterize the complex multipath propagation into a discrete set of signal taps with specific delays and average powers) provide a structured format for extracting channel data. Despite their widespread use in system evaluations, they rely on generalized statistical distributions and fail to account for the unique environmental characteristics of specific sites. Manual calibration of these models for individual sites is inefficient and lacks generalizability. Realistic site-specific channel modeling traditionally relies on Ray Tracing (RT) \cite{klautau2021generating} or empirical field measurements \cite{chizhik2023accurate,adhikari2025around,huang2020non,li2022non,guo2024characterization,yang2023dynamic,hammoud2024double,bignotte2022measurement,yusuf2021autoregressive}. Although ray tracing is widely applied, it is computationally intensive and relies heavily on precise 3D environmental databases. In addition, simulations are less capable of capturing real-world complexity and dynamics than actual measurements. However, conducting extensive field measurements is expensive, laborious, and practically unscalable in all deployment scenarios. Therefore, it is important to develop an efficient and generalizable approach for site-specific channel prediction.

To overcome the limitations of traditional modeling and extensive field measurements, Artificial Intelligence (AI)—particularly deep learning—has demonstrated excellence in handling non-linear, complex mapping relationships \cite{He-p3AI,Huang-P3AI,Zhangyx}. In existing AI-driven channel modeling research, environmental input features generally fall into two categories: hand-crafted scalar inputs (e.g., transmitter-receiver distance and regional building density) \cite{Park2020, Adejo2022, Sotiroudis2024, Sani2025, Wu2025} and image-based inputs (e.g., satellite images, 2D environment models, and building masks) \cite{Cui2019, Alqahtani2022, Iftikhar2025, Li2025,Alrabeiah,Murakami,Zhang}. The former relies heavily on simplified statistical representations, making it difficult to accurately encapsulate abstract environmental contexts, such as the specific spatial distribution and geometric structures of physical scatterers. In contrast, for site-specific channel inference, satellite imagery emerges as an accessible and information-rich alternative. Specifically, satellite images can provide global views that cover the entire propagation path, which is crucial for capturing large-scale blockages. Furthermore, they offer high-resolution local views that capture the detailed environmental topology around the receiver, which strictly dictates the behavior of small-scale multipath effects. Using the powerful spatial feature extraction capabilities of deep learning, it becomes feasible to directly extract the underlying physical propagation mechanisms from these satellite images.

Currently, models that use satellite images as input focus primarily on large-scale channel parameters (e.g., path loss and signal quality parameters) \cite{Qiu-PL,Zhu-PL,Wang-PL,Wang-PL2,wang-PL3,Kwon-PL,Xu-PL,Sani-PL,Sousa-PL-baselin,Jaensch-PL,Thrane-PL,Thrane-RSRP,Hayashi-PL}. Among these, advancing beyond approaches that rely solely on fundamental Computer Vision (CV) pipelines, several studies have incorporated physically interpretable feature designs to enhance the model's capacity to learn underlying propagation mechanisms. For example, Wang et al. developed a hybrid model-assisted framework that infuses domain-specific physical mechanisms directly into the data-driven architecture, enhancing the physical consistency and generalization capabilities of the predictions in suburban scenarios \cite{wang-PL3}. Hayashi et al. incorporated deterministic path profiles along with 2D spatial information, representing the physical line-of-sight(LOS) and terrain obstacles between transceivers to improve the network's interpretation of terrain-induced propagation losses \cite{Hayashi-PL}. Zhu et al. utilized an adaptive Meta Mask R-CNN model to identify and extract physical scatterer features from the input, bridging the gap between macroscopic environmental structures and path loss \cite{Zhu-PL}. Despite these methodological advances in incorporating physical propagation mechanisms, a critical limitation persists in the current literature that these satellite-based approaches remain confined to the prediction of large-scale fading parameters. Although some recent efforts have expanded their scope, for example, Li et al. predicted delay spread using the Fresnel zone theory \cite{Li-PL/DS}. However, these outputs are restricted to statistical parameters. Consequently, these models lack the capacity to reconstruct the complete CIR, leaving a gap in fully characterizing the multipath behaviors required for modern wireless system evaluation.

Bridging this gap in order to predict site-specific channel parameters from the satellite image requires overcoming several challenges. First, in the spatial dimension, multipath propagation is physically affected by both the macroscopic global environment and the microscopic local features. A single feature extraction pipeline is fundamentally inadequate to balance both dimensions concurrently. Second, in the temporal dimension, the dynamic birth-death processes and the evolution of multipath components (MPCs) typically exhibit a pronounced long-term memory effect. Relying solely on static, single-frame satellite images makes it difficult to track these trajectories, as capturing their continuous evolution fundamentally requires a broader temporal receptive field. Finally, regarding optimization, directly optimizing all network parameters in an end-to-end manner is challenging. Different channel parameters exhibit distinct physical sensitivities and evolutionary time scales. Forcing a neural network to learn these diverse physical characteristics simultaneously easily leads to severe gradient conflicts.

To address these challenges, we propose a deep learning based site-specific channel modeling and inference framework using satellite images. The parameters of the TDL model serve as the predictive targets, as their structured representation is compatible with neural network optimization paradigms. The framework integrates a dual-branch pipeline with a cross-attention fusion mechanism to capture how global macroscopic environments physically modulate local microscopic scattering features. Furthermore, we introduce a specialized MPCs tracking module to model the dynamic temporal evolution of tap parameters, which is optimized via a decoupled multi-stage training strategy to prevent gradient conflicts. The main contributions and innovations of this work are summarized as follows:

\begin{itemize}
  \item A dataset consisting of empirical channel measurements and satellite images is constructed. This dataset covers a variety of environments, including urban, suburban, and mixed scenarios.
  
  \item A data preprocessing pipeline is developed to extract TDL parameters from raw PDPs using dynamic thresholding and iterative peak extraction. Semantic segmentation is applied to satellite images to generate building masks.
  
  \item A novel deep learning framework, designed to fuse global and local environmental visual features, is introduced for the accurate prediction of TDL parameters.
\end{itemize}

The remainder of this paper is organized as follows: Section II outlines the proposed framework. Section III presents the channel measurement. Section IV details the data pre-processing. Section V details the architecture of the proposed model. Section VI evaluates the performance of the model. Finally, Section VII concludes the paper.

\section{Channel Inference Framework}
\begin{figure}[t]
    \centering
    \includegraphics[width=\linewidth]{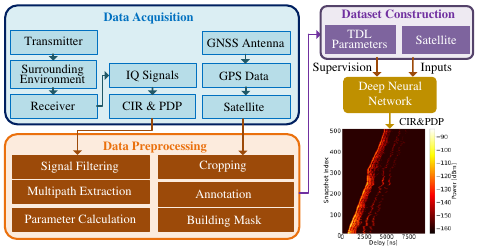}
    \caption{An overview of the proposed framework.}
    \label{fig:system}
\end{figure}

We propose an AI-based site-specific channel modeling and inference approach using satellite images, with the overall framework illustrated in Fig. \ref{fig:system}. The proposed method uses both channel data and satellite data to train a model capable of predicting CIR. Channel data can typically be obtained through Ray-Tracing simulations or real-world measurements. Compared to Ray-Tracing, measurement-based data offer greater authenticity and representativeness in capturing environmental complexity and dynamic variations, making them more suitable for evaluating the generalization capability of the model. We built a measurement platform capable of data acquisition.The receiver antenna acquires the raw in-phase and quadrature (IQ) signals, which are calibrated and transformed to obtain the CIR and PDP. Meanwhile, the GPS data collected by the Global Navigation Satellite System (GNSS) antenna are used to extract the corresponding site-specific satellite image.

During the data preprocessing stage, the TDL parameters are chosen to abstract the raw channel information, as their highly structured format is compatible with neural network training. The raw PDP undergos a refinement comprising three sequential operations: signal filtering to denoise, multipath extraction to identify dominant propagation paths, and parameter calculation to quantify the statistical channel behavior. Specifically, the TDL parameters comprise the large-scale fading parameter $\mathcal{P}$ (characterized by the first tap power), the K-factor $K$, the total number of taps $N$, the tap Normalized delays $ \tau=\left\{ \tau_1, \tau _2, \cdots \tau _N\right\}$, and their corresponding tap powers$ p=\left\{ p_1, p_2, \cdots p_N\right\}$. It is worth noting that as this work is conducted under LOS scenarios, the K-factor $K$ practically corresponds to the first tap, which follows a Rician distribution. The TDL parameters can be expressed as:
\begin{equation}
\Theta=\left\{\mathcal{P}, K, N, \tau, p \right\}
\end{equation}
\begin{figure}[t]
    \centering
    \includegraphics[width=\linewidth]{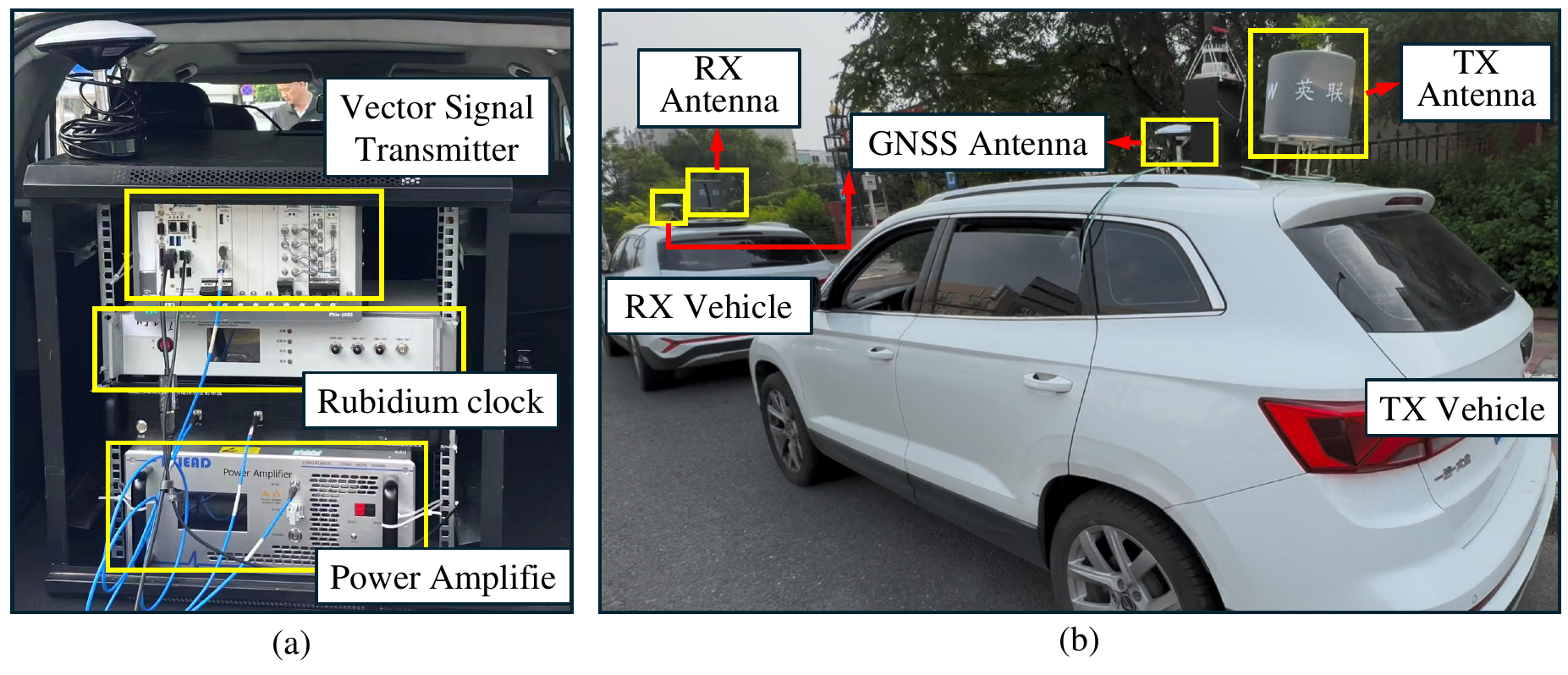}
    \caption{Measurement system architecture and key equipment: (a) TX vehicle equipment, (b) measurement vehicles and antennas.}
    \label{fig:measurement}
\end{figure}
\begin{table}[t]
\centering
\caption{Configurations of measurement system}
\begin{tabular}{lc}
\toprule
\textbf{Parameters} & \textbf{Value} \\
\midrule
Center frequency           & 5.9\,GHz \\
Bandwidth                  & 30\,MHz \\
Transmit power             & 45\,dBm \\
Sounding signal            & Multi-carrier signal \\
Number of frequency points & 1024 \\
Transmitter antenna        & Omnidirectional antenna \\
Receiver antenna           & Omnidirectional antenna \\

\bottomrule
\end{tabular}
\label{tab:config}
\end{table}

The collected satellite imagery is preprocessed through image cropping , annotation, and mask generation operations. These image processing steps establish a satellite dataset comprising global satellite views that cover the entire propagation path, local satellite views capturing environmental details strictly around the RX, and binary building masks representing environmental scatterers and blockages.

Subsequently, the processed satellite image $S$ and the set of TDL parameters $\Theta$ are paired sample by sample to construct the dataset based on timestamp and spatial location to ensure environmental consistency across the training data. Given that both the channel measurements and the GPS data are time-stamped by the same rubidium clock, their alignment is inherently achieved without manual effort. The model is trained on the constructed dataset to learn a mapping from the satellite image to the TDL parameters:
\begin{equation}
\hat{\Theta}=f(S)
\end{equation}
where $S$ denotes the input set of satellite image, $\hat{\Theta}$ is the predicted TDL parameters, and $f$ represents the proposed model, which will be elaborated in Section V.

\section{Channel Measurements}
In this section, we describe the previous measurement campaigns conducted to validate the proposed reconstruction framework. These measurements provide the foundation for establishing the joint channel-satellite dataset, ensuring that the subsequent deep learning model is trained on physically accurate propagation characteristics.
\subsection{Measurement Setup}

As shown in Fig. \ref{fig:measurement}, the empirical data collection was carried out using a vehicle-integrated measurement platform. This setup comprises dedicated transmitting (Tx) and receiving (Rx) subsystems, built around an NI PXIe-5673 vector signal generator (VSG) and an NI PXIe-5663 vector signal analyzer (VSA), respectively. For the sounding procedure, we transmitted a 30 MHz broadband multi-carrier signal formulated with 513 subcarriers.

Regarding the radio frequency (RF) front-end, identical omnidirectional single-element antennas were deployed at both the Tx and Rx sides. Both antennas were secured on top of the test vehicles at an elevation of roughly 1.8 m. To ensure rigorous time alignment between the distributed nodes, GNSS-disciplined rubidium clocks were implemented. Crucially, these clocks continuously logged precise geographical coordinates, facilitating exact Tx/Rx localization and enabling the subsequent acquisition of corresponding site-specific satellite imagery.

A breakdown of the system parameters is provided in Table \ref{tab:config}. During the dynamic measurements, the system operated with the aforementioned 30 MHz bandwidth and continuously recorded channel snapshots at a rate of 45 frames per second.
\subsection{Measurement Campaign}
\begin{figure}[t]
    \centering
    \includegraphics[width=\linewidth]{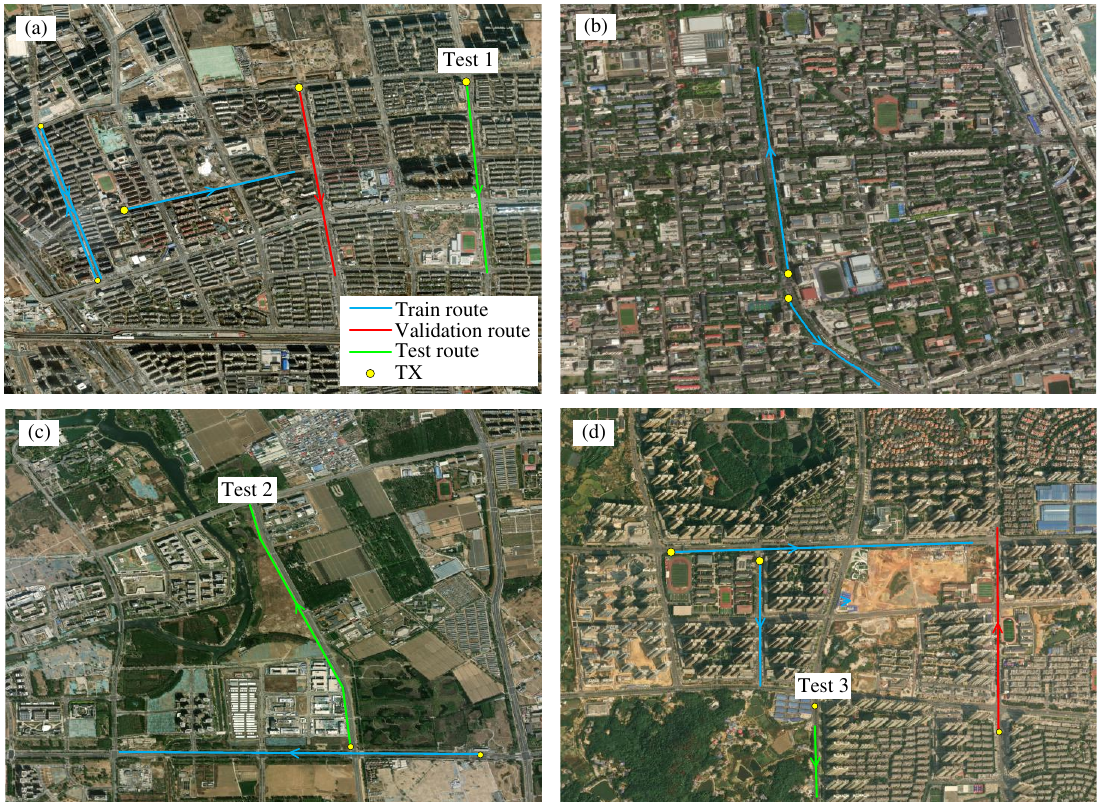}
    \caption{Measurement Scenarios and data partitioning.}
    \label{fig:route}
\end{figure}
As illustrated in Fig. \ref{fig:route}, measurement campaigns were conducted across diverse environments to establish a robust and representative dataset. The measurement locations span two cities. Specifically, the aerial maps in Fig. \ref{fig:route}(a)-(c) present the various measurement sites deployed in Beijing, china, while Fig. 3(d) presents the measurement area in Changsha, china. Furthermore, the measurement sites cover diverse environmental complexities, spanning urban, suburban, and mixed propagation scenarios. As shown in Fig. \ref{fig:route}(a) and 3(b), the typical urban areas are characterized by dense buildings, complex intersections, and rich scatterers. In contrast, Fig. \ref{fig:route}(c) illustrates a typical suburban environment with a more open layout, sparse structures, diverse vegetation, and unique spatial geometries. Finally, Fig. \ref{fig:route}(d) presents a mixed scenario containing both buildings and vegetation. This measurement campaign ensures the richness and diversity of the collected dataset.
\section{Data Pre-Processing}
In this section, we presents the dataset preprocessing and construction workflow. First, we detail the extraction of reliable TDL parameters from the raw PDPs via dynamic thresholding and iterative peak extraction techniques. Next, we describe the rigorous spatial alignment, cropping, and semantic segmentation of the satellite images. Finally, the dataset splitting methodology is summarized.
\begin{figure*}[t]
    \centering
    \includegraphics[width=\linewidth]{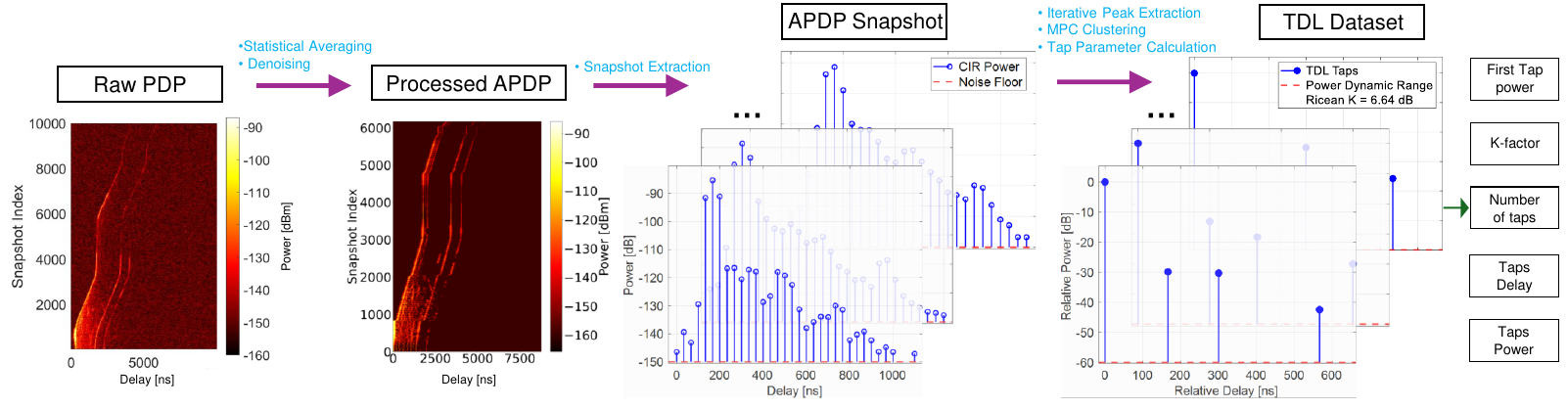}
    \caption{Procedure of constructing TDL parameters dataset.}
    \label{fig:datapreprocess}
\end{figure*}

\subsection{TDL Parameters Calculation}
Fig. \ref{fig:datapreprocess} details the construction procedure of the TDL parameter dataset. To suppress random fast fading and system noise for reliable small-scale multipath extraction, the raw PDPs undergo sliding-window statistical averaging in the time domain to yield the Averaged Power Delay Profile (APDP). Next, an adaptive dynamic thresholding method is employed for denoising. Specifically, valid APDP samples above the absolute noise floor (-160 dB) are isolated and sorted by power. The mean of the bottom 20\% samples is calculated as the real-time noise floor estimate for the given window. By adding an empirical 11 dB noise margin to this estimate, a dynamic truncation threshold is dynamically generated for each snapshot. Components falling below this threshold are classified as background noise and truncated to -200 dB. This procedure yields the fully denoised APDP, denoted as $P_{denoised}$ with length $L$, ensuring that the processed signal is strictly composed of multipath components.

Subsequently, for each snapshot, an iterative peak extraction algorithm is executed on $P_{denoised}$, as outlined in Algorithm 1. Specifically, all local power maxima are first identified as candidate taps. The algorithm then enters an iterative loop to sequentially find the maximum peak in the residual signal. To accommodate the dynamic SNR of different snapshots, a relative range threshold ($D_{th}$) is set 50 dB below the global maximum peak of $P_{denoised}$; peaks falling below this level are treated as noise, terminating the loop. Once a valid tap delay is extracted, the residual signal within a spatial masking neighborhood of $\delta = 3$ samples on either side of its center is forced to zero. This critical step prevents the energy leakage induced by channel dispersion from being falsely detected as separate multipath arrivals, ensuring that the extracted taps remain physically independent.

Once the delay positions of all multipath taps are iteratively extracted, they are sorted in ascending order of their delays. A local delay window is then defined for each tap, starting from its current delay position and terminating just before the arrival time of the subsequent tap. The tap power is computed by integrating the signal energy over this interval. After calculating the power $\mathcal{P_0}$ and the delay of the first tap, a normalization process is applied to all taps to obtain the normalized delay array $\tau$ and the power array $p$. Furthermore, the absolute maximum power of the first tap on the APDP is treated as the pure LOS path power $p_{LOS}$. The NLOS power $p_{NLOS}$, contributed by micro-scatterers arriving closely behind the main path, is then obtained by subtracting $p_{LOS}$ from the total integrated power of the first tap's local window. Finally, the Rician K-factor $K$ for the snapshot is calculated as $K = p_{LOS} - p_{NLOS}$.
\begin{algorithm}[t]
\caption{Iterative Peak Search and Tap Energy Integration}
\KwData{Denoised APDP $P_{denoised}$ of length $L$, Range threshold $D_{th}$, Min separation $\delta$}
\KwResult{Set of extracted taps $\mathcal{T} = \{(\tau_i, p_i)\}$}

\textbf{Initialize:}\\
$\mathcal{D} \gets \emptyset$, $\mathcal{T} \gets \emptyset$, $P_{res} \gets P_{denoised}$\;
$P_{limit} \gets \max(P_{denoised}) - D_{th}$\; 
Compute mask $M \in \{0,1\}^L$, where $M[\tau]=1$ iff $P_{res}[\tau]$ is a local peak\;

\tcp{Phase 1: Iterative Delay Extraction}
\While{\textbf{true}}{
    $\tau_{peak} \gets \arg\max_{\tau} (P_{res} \odot M)$ \;
    $P_{peak} \gets P_{res}(\tau_{peak})$\;
    
    \If{$P_{peak} < P_{limit}$ \textbf{or} $P_{peak} == 0$}{
        \textbf{break}\;
    }
    
    $\mathcal{D} \gets \mathcal{D} \cup \{\tau_{peak}\}$\;
    
    $\mathcal{N} \gets [\max(1, \tau_{peak} - \delta), \min(L, \tau_{peak} + \delta)]$\;
    $P_{res}[\mathcal{N}] \gets 0$, $M[\mathcal{N}] \gets 0$\;
}

\tcp{Phase 2: Delay Sorting and Power Integration}
Sort $\mathcal{D}$ ascendingly to $\{\tau_1, \tau_2, \dots, \tau_{|\mathcal{D}|}\}$\;
Set $\tau_{|\mathcal{D}|+1} \gets L + 1$ \;

\For{$i \gets 1$ \KwTo $|\mathcal{D}|$}{
    $p_i \gets \sum_{n=\tau_i}^{\tau_{i+1}-1} P_{denoised}[n]$ \;
    $\mathcal{T} \gets \mathcal{T} \cup \{(\tau_i, p_i)\}$\;
}

\Return $\mathcal{T}$\;
\end{algorithm}
\subsection{Satellite Image Pre-processing}
\begin{figure}[t]
    \centering
    \includegraphics[width=\linewidth]{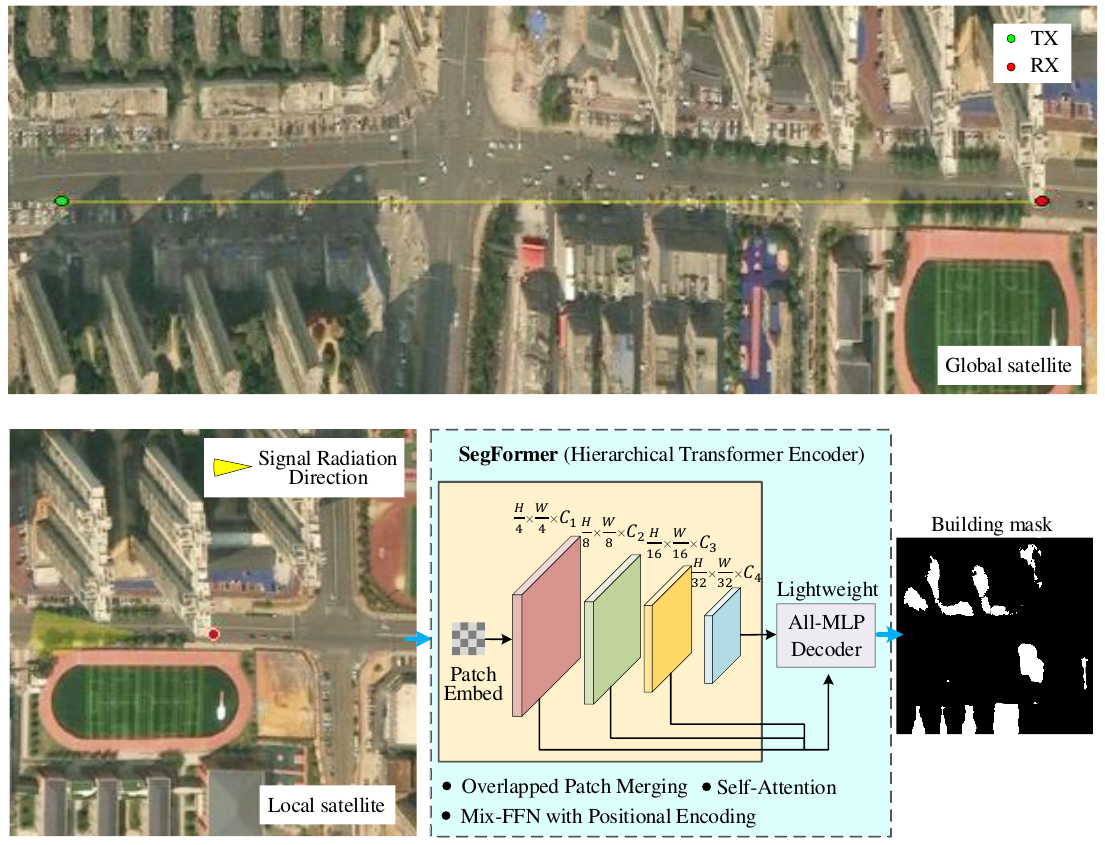}
    \caption{Procedure of construct satellite dataset.}
    \label{fig:satellite}
\end{figure}

Fig. \ref{fig:satellite} illustrates the procedure for constructing the satellite dataset. First, to capture the global large-scale blockages and environmental features between the Tx and the Rx, the physical distance and azimuth angle of the propagation link are computed based on their respective coordinates. An encompassing Region of Interest (ROI) is then cropped from the original image. Subsequently, a rotation alignment operation is performed to map all Tx-Rx links to a horizontal orientation. Furthermore, to prevent dimensional distortion caused by excessively short propagation distances, the cropping width is dynamically set to the maximum of 1.1 times the physical distance and 256 m, paired with a fixed street-level height of 128 m. The resulting image is then resized to 512 × 256 pixels. Additionally, Tx/Rx node markers and the LOS trajectory are annotated onto the image to enhance the neural network's perception of the link topology.

The local scattering environment surrounding the Rx plays a critical role in determining multipath effects. After applying the same rotation alignment, a square physical receptive field of 256 m × 256 m, centered at the Rx coordinates, is cropped. This region is then resized to a standard of 224 × 224 pixels, matching the input requirements of popular vision backbones (e.g., ResNet or ViT). Furthermore, the Rx location and the approximate direction of signal arrival are annotated onto the image.

Meanwhile, to assist the model in learning the critical physical propagation mechanism of building blockages, an advanced semantic segmentation model based on the Transformer architecture (Segformer-B1) is introduced \cite{SegFormer}. Given the relatively low resolution of the local micro-views (224 × 224), an image interpolation enhancement strategy is employed during the inference stage. Specifically, the input images are upsampled to 1024 × 1024 to fully activate the Segformer's pre-trained weights derived from high-resolution remote sensing datasets, thereby precisely capturing the edges of building footprints. After the model output is restored to its original size using bilinear interpolation, the 'building' category is extracted to generate a binary semantic mask. This mask serves as an additional modal input, guiding the network to learn pure physical scattering mechanisms.
\subsection{Dataset Construction}
The processed satellite images and the TDL parameter sets are paired on a sample-by-sample basis according to their specific timestamps and spatial locations, thereby ensuring the environmental consistency of the dataset. Given the high similarity in spatial environmental features among adjacent samples along the same route, a route-based dataset splitting strategy is adopted instead of the conventional random splitting approach. This design prevents potential spatial data leakage during the model training phase. The partitioning of the dataset is illustrated in Fig. \ref{fig:route}, comprising a total of 16,800 samples.
\section{Model Design}
\begin{figure*}
    \centering
    \includegraphics[width=\linewidth]{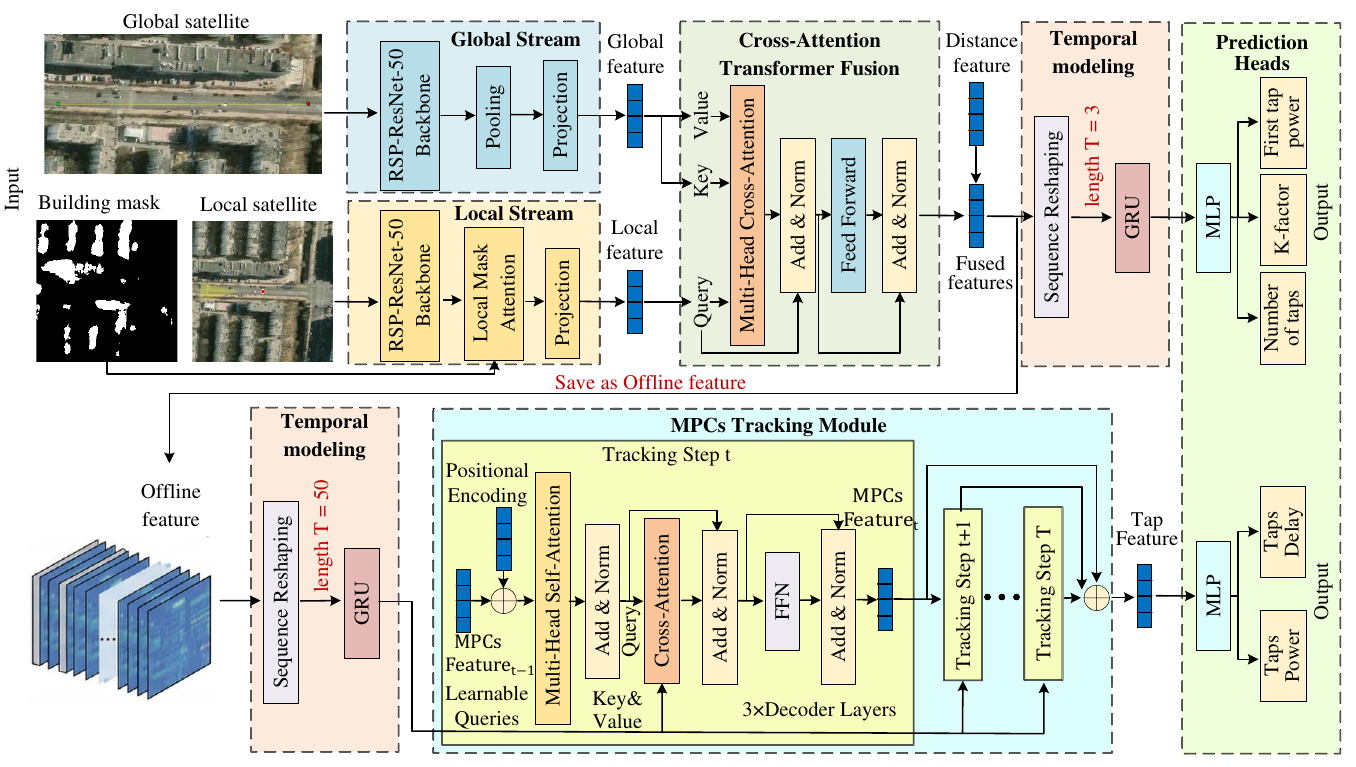}
    \caption{The proposed model.}
    \label{fig:model}
\end{figure*}
\begin{table}[htbp]
\centering
\caption{Detailed Architecture of the Proposed Model}
\label{tab:architecture}
\footnotesize 
\setlength{\tabcolsep}{2pt} 
\begin{tabular}{@{}llcl@{}}
\toprule
\textbf{Network} & \textbf{Layer} & \textbf{Output} & \textbf{Description} \\
\midrule
\multirow{3}{*}{Input} 
 & G/L Satellite & ($B \times T$, 3, $H$, $W$) & / \\
 & Mask & ($B \times T$, 1, $H$, $W$) & / \\
 & Dist. Feat. & ($B \times T$, 1) & / \\
\midrule
\multirow{3}{*}{Global} 
 & ResNet-50 & ($B \times T$, 2048, $h$, $w$) & / \\
 & Pooling & ($B \times T$, 2048) & GAP \\
 & Projection & ($B \times T$, 512) & 2048 $\to$ 512 \\
\midrule
\multirow{3}{*}{Local} 
 & ResNet-50 & ($B \times T$, 2048, $h$, $w$) & / \\
 & Mask Attn. & ($B \times T$, 2048) & GAP \\
 & Projection & ($B \times T$, 512) & 2048 $\to$ 512 \\
\midrule
\multirow{2}{*}{Cross-Attn.} 
 & MH Attn. & ($B \times T$, 512) & L:2, H:4 \\
 & Concat. & ($B \times T$, 513) & 512 + 1 $\to$ 513 \\
\midrule
\multirow{2}{*}{Temporal} 
 & Reshape & ($B$, $T$, 513) & / \\
 & GRU & ($B$, $T$, 513) & L:1, Hid:513 \\
\midrule
\multirow{2}{*}{Tracking} 
 & Pos. Enc. & ($B$, $T$, $N$, 513) & / \\
 & Decoder & ($B$, $T$, $N$, 513) & L:3, H:9 \\
\midrule
\multirow{3}{*}{Pred. Heads} 
 & MLP ($\mathcal{P}$) & ($B$, $T$, 1) & 513 $\to$ 128 $\to$ 64 $\to$ 1 \\
 & MLP ($K$/$N$) & ($B$, $T$, 1) & 513 $\to$ 128 $\to$ 1 \\
 & MLP ($\tau$/$p$) & ($B$, $T$, $N$) & 513 $\to$ 1 \\
\bottomrule
\multicolumn{4}{@{}l@{}}{\textit{Note:} G/L: Global/Local, Attn.: Attention,} \\
\multicolumn{4}{@{}l@{}}{GAP: Global Average Pooling, L: Layers, H: Heads,} \\
\multicolumn{4}{@{}l@{}}{Hid: Hidden size, $\mathcal{P}$: First tap power, $K$: K-factor,} \\
\multicolumn{4}{@{}l@{}}{$N$: Number of taps, $\tau$: Taps Delay, $p$: Taps Power.}
\end{tabular}
\end{table}
In this section, we detail the proposed model framework for inference of site-specific channel parameters. The model integrates a Cross-Attention Transformer Fusion module that dynamically modulates local spatial features using global environmental constraints. The model incorporates specialized temporal prediction modules. These include a short-sequence temporal modeling component and a long-sequence MPCs tracking module designed to rigorously capture the dynamic birth-death processes of tap parameters. We subsequently outline a decoupled, multi-stage training strategy.
\subsection{Model Structure}
The proposed model aims to process satellite images, learn macro- and micro-propagation mechanisms in wireless channels, and extract spatiotemporal features for predicting TDL parameters, as shown in Fig. \ref{fig:model}. The framework employs a dual-branch feature extraction pipeline. To capture large-scale fading effects, the Global Stream processes global satellite images using a Remote Sensing Pretraining-ResNet-50 (RSP-ResNet-50) backbone to extract the global features of the LOS link. Meanwhile, since small-scale fading is primarily dependent on the environment surrounding the RX, the Local Stream is tasked with extracting local features. Moreover, a Local Mask Attention mechanism is introduced to fuse the building mask with these local features. This design compels the network to focus on significant buildings responsible for signal reflection and diffraction, suppressing interference from irrelevant features and enhancing the physical rationality of the local feature extraction.

Multipath propagation is physically constrained by the global environmental, for instance, dense urban buildings in urban areas trigger multiple reflection and diffraction, multipath effects in suburban areas are relatively weak. To capture this modulating effect of the macroscopic environment on microscopic multipath phenomena, the model employs a Cross-Attention Transformer Fusion module to perform cross-scale feature interaction. Specifically, the local features representing the microscopic environment serve as the Query vectors, while the global features denoting the macroscopic environment act as the Key and Value vectors. This fusion strategy enables the global features to dynamically modulate the local ones. Acting like a filter, it empowers the model to capture which specific local features are critical to the multipath effects under the current macroscopic scenario.

\begin{figure*}[t]
    \centering
    \includegraphics[width=\linewidth]{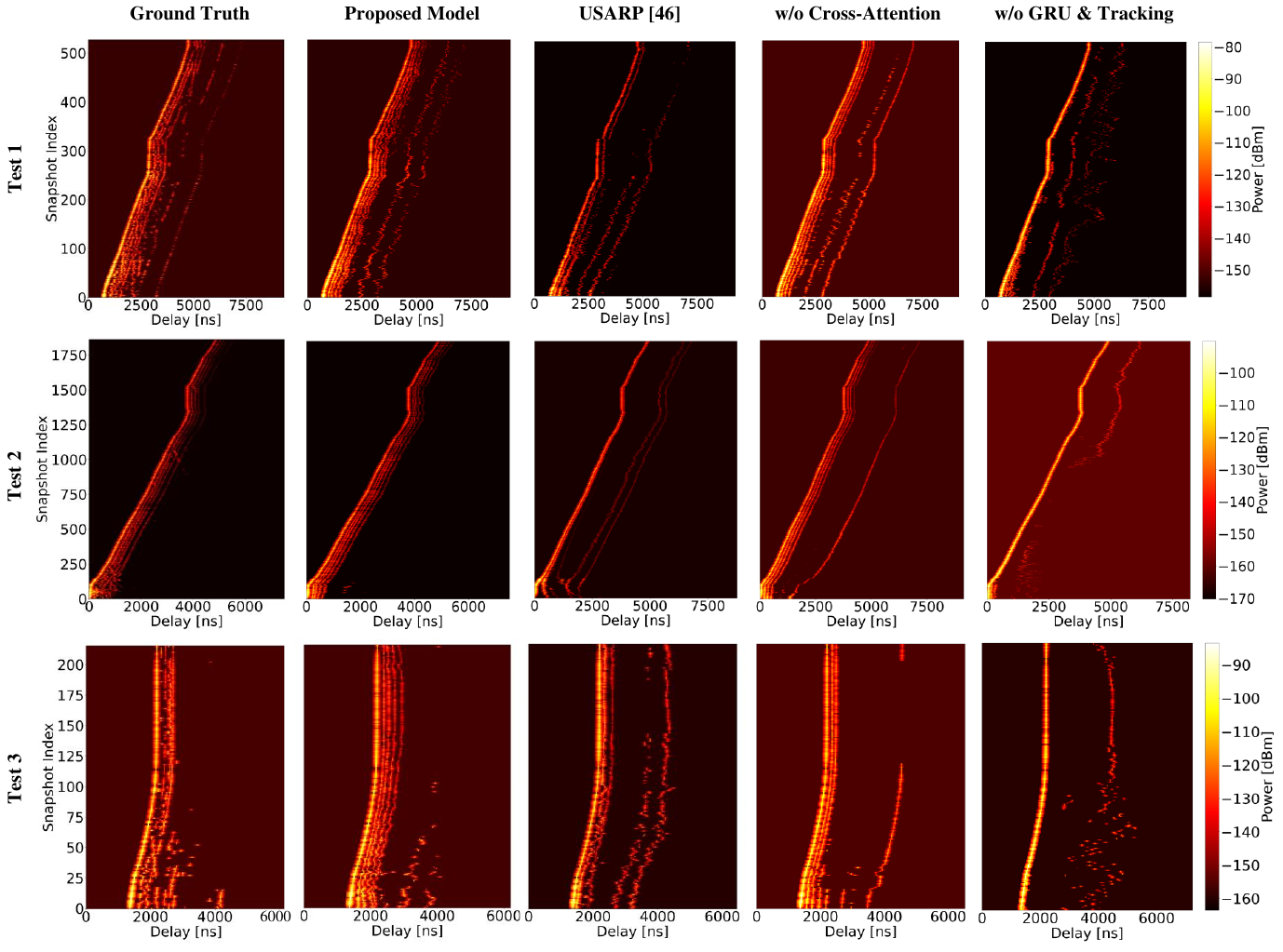}
    \caption{Comparison of predicted PDPs under different model configurations. From left to right are the PDPs of the ground truth, the proposed model, the adapted USARP baseline, the ablation variant without Cross-Attention (w/o Cross-Attention), and the ablation variant without GRU \& Tracking (w/o GRU \& Tracking).}
    \label{fig:result}
\end{figure*}
\begin{figure}[t]
    \centering
    \includegraphics[width=\linewidth]{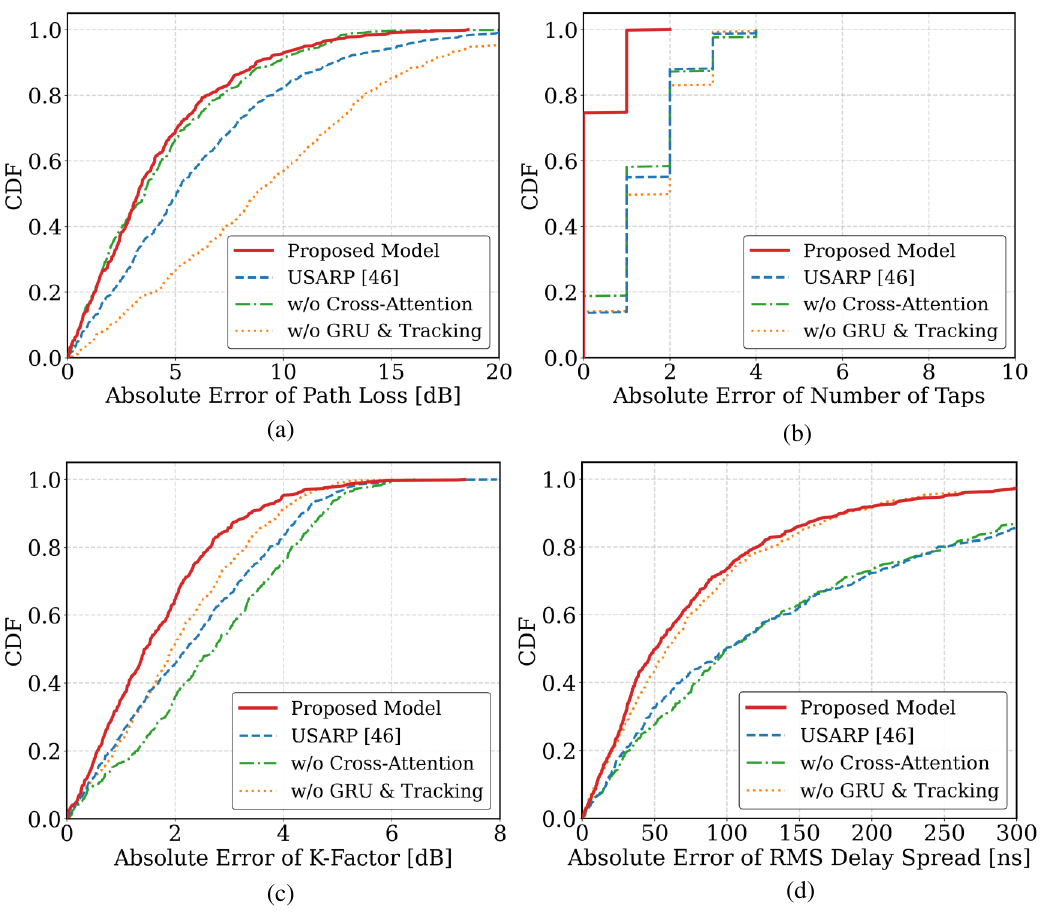}
    \caption{CDF of absolute prediction errors for different channel parameters.}
    \label{fig:AECDF}
\end{figure}

Given the gradual evolution of the physical environment along the continuous trajectory of the RX, the large-scale fading and multipath effects of the channel typically exhibit high spatial consistency and temporal correlation over short periods. To prevent observation errors or local visual abruptions in single-frame static satellite images from causing non-physical drastic fluctuations in the predicted parameters, the fused features are concatenated with the Tx-Rx distance and subsequently input into a short-sequence temporal modeling module. This module, employing a Gated Recurrent Unit (GRU) with a time step of $T_s=3$, is designed to capture the short-term spatial smoothness between adjacent snapshots along each route, thereby regularizing the feature representations. Ultimately, the output of this module is passed through a Multi-Layer Perceptron (MLP) prediction head to regress the preliminary TDL parameters, including the first tap power, K-factor, and the number of taps.

An independent MPC Tracking Module is designed to predict the complex, dynamically evolving tap delay and power arrays. The birth-death processes and evolution of MPCs typically exhibit a pronounced long-term memory effect. a broader temporal receptive field is required to capture their evolution. To achieve this while avoiding computational overhead and GPU memory bottlenecks, offline features are introduced. Specifically, the frame-level spatial fusion features output by the Cross-Attention Transformer Fusion module are cached offline. Before the Tracking Module, these offline features are directly retrieved and reorganized into long sequences ($T_l =50$), and passed to a separate GRU for long-term temporal encoding. Following this, a recurrent Transformer decoder architecture is employed. At each tracking step $t$, the prior multipath state ($\text{MPCs Feature}_{t-1}$) and a learnable query positional encoding compute multi-head self-attention, followed by cross-attention with the temporally encoded offline features. This iterative attention-based tracking models the temporal coherence and trajectories of MPCs as the RX moves. Finally, the aggregated tap features are routed to specific MLP heads to predict tap delays and powers. The detailed architecture of the proposed model is shown in Table II.

\subsection{Loss Functions and Training Strategies}
In the proposed model, directly optimizing all parameters in an end-to-end manner is a highly challenging task. The TDL parameters exhibit distinct physical sensitivities and evolutionary time scales. To prevent gradient conflicts and ensure that each module learns its designated physical mechanism, we adopt a decoupled, multi-stage training strategy. The network is trained progressively, with each stage freezing specific upstream modules to stabilize the feature representations.
\subsubsection{Training Stage I: Large-scale fading $\mathcal{P}$}:
In the first stage, the network focuses on extracting the Global feature. The Global Stream, the First Tap Power prediction head, and the GRU are activated, while all other modules are frozen. Since the first tap power represents the large-scale fading which varies smoothly along the moving trajectory, we formulate the loss function as a combination of Mean Squared Error (MSE) and a temporal consistency regularization term:
\begin{equation}
\mathcal{L}_1 = \frac{1}{B \cdot T_s} \sum_{b=1}^B \sum_{t=1}^{T_s} \left| \hat{\mathcal{P}}_{b,t} - \mathcal{P}_{b,t} \right|^2 + \lambda_1 \mathcal{L}_{\mathrm{temp}}(\hat{\mathcal{P}})
\end{equation}
where $B$ is the batch size, $T_s=3$ is the short sequence length used in this stage, and $\hat{\mathcal{P}}_{b,t}$ and $\mathcal{P}_{b,t}$ are the predicted and ground-truth normalized first tap powers, respectively. The temporal consistency loss $\mathcal{L}_{\mathrm{temp}}$ penalizes abrupt, non-physical fluctuations between adjacent snapshots to mitigate observation noise from static satellite images:
\begin{equation}
\mathcal{L}_{\mathrm{temp}}(\hat{\mathcal{P}}) = \frac{1}{B(T_s-1)} \sum_{b=1}^B \sum_{t=2}^{T_s} \left| \hat{\mathcal{P}}_{b,t} - \hat{\mathcal{P}}_{b,t-1} \right|^2
\end{equation}
where $\lambda_1$ is empirically set to 0.1.
\subsubsection{Training Stage II: Statistical Channel Attributes ($K$ and $N$)}
The second stage aims to parameterize the statistical properties of the local scattering environment, specifically the Ricean K-factor and the total number of MPC taps. The Global Stream is frozen to serve as a stable macroscopic context provider, while the Local Stream, the Cross-Attention Fusion module, and their corresponding prediction heads are optimized. The loss function for this stage is defined as:
\begin{equation}
\mathcal{L}_2 = \mathrm{MSE}(\hat{K}, K) + \mathrm{MSE}(\hat{N}, N) + \lambda_1 \mathcal{L}_{\mathrm{temp}}(\hat{K})
\end{equation}
where $\hat{K}$ and $\hat{N}$ denote the predicted K-factor and the number of taps, respectively. By isolating this training stage, the Local Stream is forced to learn how the local semantic masks (e.g., buildings) interact with the global path constraints to determine the multipath richness, preventing the network from prematurely overfitting to the exact delays of individual MPCs.
\begin{figure}[t]
    \centering
    \includegraphics[width=\linewidth]{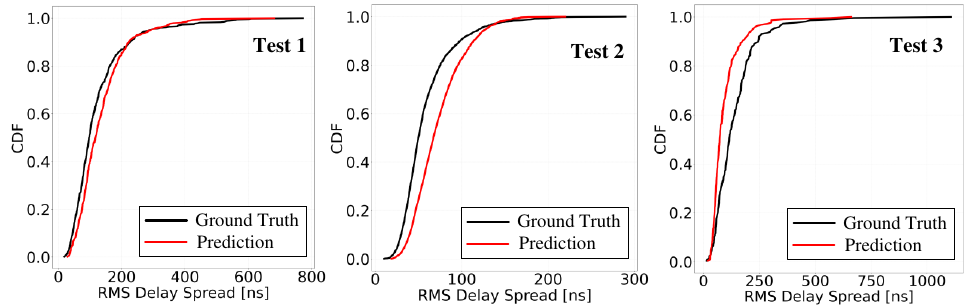}
    \caption{CDF comparison between the proposed model and measured data.}
    \label{fig:CDF}
\end{figure}

\subsubsection{Training Stage III: MPCs Tracking ($\tau$ and $p$)}
The final stage reconstructs the fine-grained TDL parameters by predicting the specific delay and relative power of each tap. Tracking the birth, evolution, and death of individual multipath clusters requires a significantly larger temporal receptive field. We extract and freeze the output representations from the previously trained spatial fusion modules and save them as offline features. This allows us to drastically expand the sequence length to $T_l =50$ for the MPC Tracking Module without repeatedly executing the heavy CNN backbones.

Since the extracted MPCs are essentially an unordered set of points, standard point-to-point loss functions cannot be directly applied. We model the MPCs prediction as a bipartite matching problem and employ a matching loss coupled with a physical repulsion constraint:
\begin{equation}
\mathcal{L}_3 = \mathcal{L}_{\mathrm{match}} + \alpha \mathcal{L}_{\mathrm{rep}}
\end{equation}
where $\alpha$ is a learnable weight parameter updated via backpropagation during training to dynamically balance the loss terms. The matching loss $\mathcal{L}_{\mathrm{match}}$ utilizes the Hungarian algorithm \cite{Hungarian} to find the optimal assignment $\sigma$ between the predicted tap set $\hat{Y} = \left\{ \hat{\tau}, \hat{p} \right\}$ and the ground-truth tap set $Y = \left\{ \tau, p \right\}$:
\begin{equation}
\mathcal{L}_{\mathrm{match}} = \frac{1}{B \cdot T_l} \sum_{b, t} \frac{1}{|V|} \sum_{j \in V} \left| \hat{y}_{\sigma(j)} - y_j \right|^2
\end{equation}
where $y_j = [w_d \cdot \tau_j, p_j]$ represents the scaled delay and power of the $j$-th ground-truth tap, $w_d$ is a weighting factor to balance the scale difference between the delay and power domains, and $V$ is the set of valid (non-padded) ground-truth taps.

To prevent the Transformer decoder from collapsing multiple predicted queries into the same dominant multipath cluster, we introduce a repulsion loss $\mathcal{L}_{\mathrm{rep}}$ to enforce a minimum physical separation threshold $d_{\mathrm{min}}$ between any two predicted delays:
\begin{equation}
\mathcal{L}_{\mathrm{rep}} = \frac{1}{B \cdot T_l \cdot N(N-1)} \sum_{b,t} \sum_{i \neq j} \max(0, d_{\mathrm{min}} - |\hat{\tau}_i - \hat{\tau}_j|)
\end{equation}
where $d_{\mathrm{min}}$ corresponds to the propagation distance equivalent to the minimum delay resolution of the receiver, which is set at 33 ns.

Regarding the optimization details, we employ the Adam optimizer in all stages. The initial learning rate is set to $1 \times 10^{-4}$ for the spatial feature extraction stages (stages I and II) and is reduced to $1 \times 10^{-5}$ for the fine-grained MPCs tracking stage (stage III) and subsequent joint fine-tuning. Early stopping is triggered if the combined validation loss fails to improve over 20 consecutive epochs, ensuring optimal generalization while preventing overfitting.

\section{Valuation And Analysis}\
\begin{figure}[t]
    \centering
    \includegraphics[width=\linewidth]{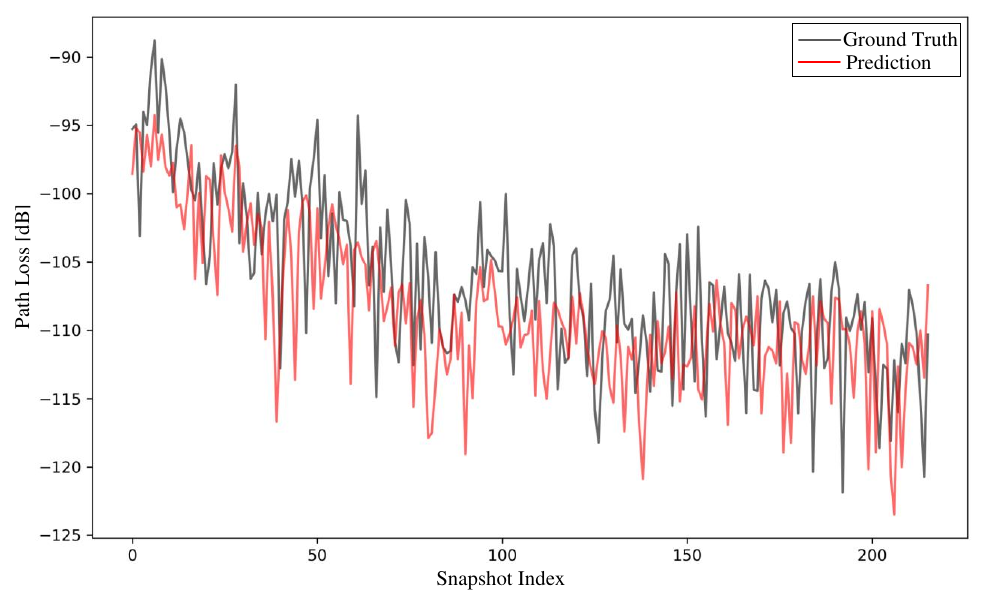}
    \caption{Path loss comparison between the proposed model and measured data.}
    \label{fig:PL}
\end{figure}

In this section, we evaluate the performance of the proposed site-specific channel inference framework. We detail the process of reconstructing the CIR from the predicted TDL parameters. Then key performance metrics are introduced. Subsequently, ablation studies and a baseline comparison are conducted to evaluate the performance of the proposed model. Finally, a complexity analysis evaluates the online inference speed in real-world wireless communication systems.

\subsection{CIR Reconstruction}
To evaluate the physical validity of the proposed model, it is necessary to reconstruct the corresponding CIR using the set of the predicted TDL parameters $\hat\Theta=\left\{\hat{\mathcal{P}}, \hat{K}, \hat{N}, \hat{\tau}, \hat{p} \right\}$ and simulate the discrete sampling at the RX. The purpose of this discrete sampling step is to emulate the hardware behavior of real-world signals arriving at the RX. By aligning the signal processing pipeline with the actual channel sounding operations, we ensure consistency between the reconstructed CIR and the measurement CIR, thereby guaranteeing the reliability of the model evaluation. For the first tap that encompasses both the LOS and the NLOS components, their power can be partitioned according to the predicted K-factor $\hat{k}$:
\begin{equation}
\hat{p}_{\mathrm{LOS}} = \hat{\mathcal{P}} \frac{\hat{k}}{\hat{k} + 1}, \quad \hat{p}_{\mathrm{NLOS}} = \hat{\mathcal{P}} \frac{1}{\hat{k} + 1}
\end{equation}
The complex coefficient for the first tap is then generated by combining a deterministic LOS component with a random Gaussian scattering component:
\begin{equation}
c_0 = \sqrt{\hat{p}_{\mathrm{LOS}}} e^{j 2\pi \theta} + \sigma_0 (x_0 + j y_0)
\end{equation}
where $\theta \sim \mathcal{U}(0,1)$ is a random phase, $x_0, y_0 \sim \mathcal{N}(0,1)$ are standard normal variables, and $\sigma_0 = \sqrt{\hat{p}_{\mathrm{NLOS}} / 2}$. For all subsequent late-arriving taps ($i > 0$), which are assumed to be Rayleigh-faded, the complex coefficients are generated as:
\begin{equation}
c_i = \sigma_i (x_i + j y_i), \quad \sigma_i = \sqrt{\hat{p}_i / 2}
\end{equation}
Furthermore, since the dataset is constructed from long-route LOS scenarios data, the absolute delay of the first tap, $\tilde{\tau}_0$, can be simulated using the physical Time of Flight (ToF), which is computed by dividing the transceiver distance by the speed of light. The absolute delays of all subsequent taps are then determined by adding their predicted taps delays to ToF, i.e., $\tilde{\tau}_i = \tilde{\tau}_0 + \hat{\tau}_i$. After obtaining the set of complex coefficients $\{c_i\}$ and their corresponding absolute delays $\{\tilde{\tau}_i\}$, we reconstruct the discrete-time baseband CIR over a uniform sampling grid $t_n = n T_s$, where $T_s = 1/f_s$ is the system sampling period. To account for the band-limited nature of the measurement system and the energy smearing caused by pulse shaping, we convolve the ideal Dirac impulses with a sinc function windowed by a Hamming window. The sampled CIR at time index $n$ is formulated as follows:
\begin{equation}
h[n] = \sum_i c_i \cdot \mathrm{sinc}\left( \frac{t_n - \tilde{\tau}_i}{T_s} \right) \cdot w(t_n - \tilde{\tau}_i)
\end{equation}
where $w(t)$ represents the Hamming window function applied over a truncated temporal support (e.g., $\pm 4 T_s$) to suppress spectral leakage. Finally, the PDP is obtained by taking the squared magnitude of the reconstructed CIR, $|h[n]|^2$, scaled by the predicted first-tap power $\hat{\mathcal{P}}$ to restore the macroscopic path loss.
\begin{figure}[t]
    \centering
    \includegraphics[width=\linewidth]{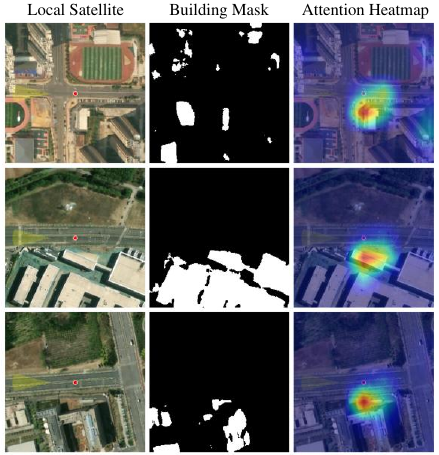}
    \caption{Visualization of attention heatmaps alongside local satellite images and corresponding building masks across different propagation scenarios.}
    \label{fig:heatmap}
\end{figure}

To assess the accuracy of the proposed model in channel inference, we employ four key performance metrics: Root Mean Square Error (RMSE) of Path Loss, RMSE of Delay Spread, RMSE of K-Factor, and PDP Average Cosine Similarity. RMSE metrics are utilized to evaluate the absolute prediction deviations of macroscopic fading, temporal dispersion, and Rician fading statistics, respectively. Among these, the RMS Delay Spread, denoted as $\tau_{\mathrm{rms}}$, is a critical parameter characterizing the time spread of the multipath channels. Based on the reconstructed discrete PDP $P[n] = |h[n]|^2$, it formulated as follows:
\begin{equation}
\tau_{\mathrm{rms}} = \sqrt{\frac{\sum_n P[n] t_n^2}{\sum_n P[n]} - \left(\frac{\sum_n P[n] t_n}{\sum_n P[n]}\right)^2}
\end{equation}
where $t_n$ represents the delay of the $n$-th sample.

Furthermore, while RMSE measures absolute scalar errors, it is equally important to assess the structural alignment and shape conformity of the multipath profiles. To this end, we introduce the Average Cosine Similarity to evaluate the structural correlation between the ground truth PDP vector $\mathbf{P}$ and the predicted PDP vector $\hat{\mathbf{P}}$. This metric inherently factors out the absolute power scale differences, focusing purely on the relative multipath distribution and structural alignment:
\begin{equation}
S_{\mathrm{cos}}(\mathbf{P}, \hat{\mathbf{P}}) = \frac{\mathbf{P} \cdot \hat{\mathbf{P}}}{|\mathbf{P}|_2 |\hat{\mathbf{P}}|_2} = \frac{\sum_n P[n] \hat{P}[n]}{\sqrt{\sum_n (P[n])^2} \sqrt{\sum_n (\hat{P}[n])^2}}
\end{equation}
where values approaching 1 indicate a remarkably high degree of structural similarity and accurate microscopic multipath reconstruction between the predicted and actual channel responses.
\begin{table*}[htbp]
\centering
\caption{Performance Comparison of Different Methods In Tests}
\label{tab:performance_comparison}
\begin{tabular}{llcccc}
\toprule
\textbf{Test Route} & \textbf{Method} & \textbf{RMSE Path Loss [dB]} & \textbf{RMSE Delay Spread [ns]} & \textbf{RMSE K-Factor [dB]} & \textbf{PDP Average Cosine Similarity} \\
\midrule
\multirow{4}{*}{Test 1} & Proposed Model & \textbf{5.62} & \textbf{97.45} & \textbf{2.46} & \textbf{0.9643} \\
                        & USARP [46] & 7.71 & 227.88 & 3.76 & 0.9071 \\
                        & w/o Cross-Attention & 8.39 & 206.86 & 3.11 & 0.9229 \\
                        & w/o GRU \& Tracking & 10.86 & 122.29 & 2.91 & 0.8942 \\
\midrule
\multirow{4}{*}{Test 2} & Proposed Model & \textbf{4.71} & \textbf{44.48} & \textbf{2.85} & \textbf{0.9735} \\
                        & USARP [46] & 7.25 & 127.60 & 3.07 & 0.9395 \\
                        & w/o Cross-Attention & 9.03 & 283.82 & 3.36 & 0.9292 \\
                        & w/o GRU \& Tracking & 11.03 & 98.43 & 4.08 & 0.9023 \\
\midrule
\multirow{4}{*}{Test 3} & Proposed Model & \textbf{5.21} & \textbf{105.53} & \textbf{1.50} & \textbf{0.9616} \\
                        & USARP [46] & 7.00 & 159.54 & 1.66 & 0.9443 \\
                        & w/o Cross-Attention & 5.53 & 189.47 & 2.01 & 0.9258 \\
                        & w/o GRU \& Tracking & 8.77 & 203.16 & 1.68 & 0.9361 \\
\bottomrule
\end{tabular}
\end{table*}
\subsection{Ablation Studies}
To validate the effectiveness of the two core modules in the proposed model (the Cross-Attention Transformer Fusion module and the temporal modeling GRU coupled with the MPCs Tracking Module) and to evaluate their actual contributions to the predictive performance, two sets of ablation studies were conducted. First, the Cross-Attention Transformer Fusion module was removed to analyze the resulting predictions. Specifically, the global features extracted by the Global Stream and the local features from the Local Stream were directly concatenated along the channel dimension, followed by a MLP for dimensionality reduction. Second, all temporal modeling GRUs and the MPCs Tracking Module were removed. Under this configuration, the model degenerates into a pure 'single-frame input and MLP output' architecture. The input sliding window data ($B \times T$) is flattened into completely independent single-frame images for forward propagation. Following feature extraction, a large MLP is employed to output all multipath delays and powers in parallel.

Fig. \ref{fig:result} illustrates the prediction results for the three routes in the test set, while Table III presents the corresponding evaluation metrics. As observed, the PDPs predicted by the model without the Cross-Attention Transformer Fusion module (w/o Cross-Attention) fail to fully capture the multipath distribution. For example, spurious multipath components emerge in the predicted PDP of Test 2 under the suburban scenario. This indicates that a simple linear concatenation is insufficient to model the complex non-linear modulation effects exerted by the macroscopic propagation environment on local microscopic scattering. Furthermore, the performance degradation across all evaluation metrics for this variant validates the necessity of the Cross-Attention Transformer Fusion module.

Similarly, the removal of the temporal modeling GRU and the MPC Tracking Module (w/o GRU \& Tracking) severely paralyzes the model's predictive stability and physical coherence. The PDPs of this variant in Fig. \ref{fig:result} exhibit a chaotic multipath distribution, where the multipath components fail to maintain continuous trajectories. This shows that predicting dynamic, highly correlated channel parameters solely from single-frame images leads to severe temporal inconsistency. Moreover, the decline in all metrics for this variant validates the necessity of the temporal memory and MPC tracking mechanisms.

Furthermore, Fig. \ref{fig:AECDF} presents the Cumulative Distribution Functions (CDFs) of the absolute prediction errors for the core parameters under different model configurations. As observed in Figs. \ref{fig:AECDF}(a)-(d), when either the cross-attention mechanism (w/o Cross-Attention) or the temporal modeling and tracking modules (w/o GRU \& Tracking) are removed, the prediction errors for all channel parameters exhibit significant degradation.
\subsection{Performance Comparison}
To evaluate the performance of the proposed model on site-specific channel inference, the Ubiquitous Satellite Aided Radio Propagation(USARP) model was selected as a baseline \cite{ Sousa-PL-baselin }. To the best of our knowledge, no existing deep learning method predicts the full CIR relying solely on satellite images. We chose USARP due to its structural and motivational similarities to our framework, as both employ satellites and a ResNet backbone. Because the original USARP targets a single scalar regression (path loss), we implemented targeted architectural and training adaptations to ensure a fair multi-task comparison.

First, since USARP cannot handle multi-view inputs or building masks, we utilized only the global satellite images. To preserve USARP's core innovation, a hard-attention binary ROI mask was applied to the global satellites generated from Tx/Rx coordinates and their LOS link. The physical distance vectors were directly assigned to the required USARP's format $\log_{10}(d_{3D})$. Second, to accommodate sliding window sequences, the input tensors $(B, T, C, H, W)$ were flattened into $(B \times T, C, H, W)$ during the forward pass for independent single-frame extraction, with dimensions restored post-prediction. Moreover, USARP's original output layer (SLP 4) forces a linear addition of image and physical features $PL = f_1(\text{image}) + f_2(\text{physical})$. Recognizing that this linear addition lacks physical rationality for complex sequence and categorical predictions, SLP 4 was upgraded into a multi-head structure. The 'linear physics injection' is retained strictly for the First Tap Power prediction, while adding parallel Fully Connected (FC) heads to predict the K-factor, tap count, delays, and powers via feature concatenation. Finally, a two-stage training strategy was applied to the baseline: initially freezing the ResNet backbone to train the new multi-head layers, followed by a global unfreezing and joint fine-tuning at a lower learning rate to guaranty optimal convergence.

As demonstrated in Table III, the proposed model significantly outperforms the baseline USARP method in all evaluation metrics on all unseen test routes. The results in Fig. \ref{fig:result} further corroborate the superiority of our framework. Although the adapted USARP model can approximately simulate the fundamental LOS path and adjacent multipath components, it fails to predict the distribution of late-arriving multipath. Consequently, this leads to a severely truncated delay profile and an inaccurate structural representation. This deficiency demonstrates that relying exclusively on global satellite imagery features is inadequate for precise site-specific channel inference. Furthermore, it firmly validates the advantages of the core modules in the proposed model for characterizing the complete multidimensional properties of wireless channels. 

Fig. \ref{fig:AECDF} shows that the proposed model exhibits superiority across all four metrics, with its error curves approaching a probability of 1 at a considerably faster rate. In particular, although the USARP baseline manages to maintain a rudimentary predictive trend for path loss, it demonstrates severe prediction deviations for the RMS delay spread and the K-factor, which characterize the microscopic multipath structure. Its error distributions are widely spread and struggle to converge. In contrast, the proposed model strictly confines the vast majority of absolute prediction errors to low levels. This fully exposes the fundamental limitations of the baseline method—which relies solely on global satellite imagery for scalar regression. This also validates the superiority of the proposed dual-stream spatial fusion combined with the multipath tracking architecture in channel reconstruction tasks. Other dimensions metrics were also used to validate the performance of the proposed model. Figs \ref{fig:CDF} and \ref{fig:PL} show, respectively, the comparison of the RMS delay spread CDF and the path loss calculated using the reconstructed CIR in the test set, demonstrating the model’s performance.

To further illustrate the physical interpretability of the proposed model, Fig. \ref{fig:heatmap} visualizes the attention heatmaps generated by the network in different scenarios. Fig. \ref{fig:heatmap} presents the original local satellite images, the binary building masks extracted through semantic segmentation, and the corresponding heatmaps overlaid with the network's attention weights. As observed, the proposed model exhibits a clear focus on the physical scatterers. In dense urban scenarios (top row), the attention is concentrated on building edges and intersections, which are the primary sources of signal reflection and diffraction. In suburban environments (middle and bottom rows), the attention dynamically shifts to capture isolated structures and specific geometric blockages. This visual evidence confirms that, driven by the Local Mask Attention and Cross-Attention Transformer Fusion modules, the network successfully learns to localize the essential macroscopic and microscopic scatterers that dictate the underlying multipath propagation mechanisms.
\subsection{Complexity Analysis}
To evaluate the feasibility of deploying the proposed model for practical site-specific channel modeling, the inference pipeline is decoupled into offline satellite image processing and online channel prediction to analyze its computational complexity and inference speed. During the offline phase, extracting multi-scale satellite images and semantic masks incurs a computational complexity of approximately 63.6 GFLOPs per sample, with an average processing time of 190 ms per sample. In the online inference phase, the model leverages these offline features to dynamically track and predict channel parameters, requiring a computational load of 25.7 GFLOPs. Evaluated on a hardware platform equipped with an Intel Core Ultra 9 275HX CPU and an NVIDIA RTX 5080 GPU, this translates to an inference time of 7 ms per sequence. Even on general-purpose computing platforms lacking GPU acceleration, the lightweight nature of the online temporal tracking ensures highly efficient execution, thereby corroborating the proposed model's potential for deployment in real-world communication scenarios.
\section{CONCLUSION}
In this paper, we proposed a deep learning based site-specific channel modeling and inference framework using satellite images to predict structured TDL parameters. We first established a joint channel-satellite dataset derived from empirical measurements in various propagation environments. Within our proposed model, a cross-attention-fused dual-branch pipeline extracts macroscopic and microscopic environmental features, while a recurrent tracking module captures the long-term dynamic evolution of multipath components. Experimental evaluations demonstrated that our framework achieves high-quality reconstruction of the complete CIR in unseen scenarios. Complexity analysis confirms the high efficiency of the decoupled inference pipeline, which requires an online inference time of 7 ms per sequence. This work provides a highly efficient and scalable pathway toward site-specific channel inference for future dynamic wireless networks.
\bibliographystyle{IEEEtran}
\bibliography{text}

\ifCLASSOPTIONcaptionsoff
  \newpage
\fi

 \end{document}